\documentclass[prx,aps,twocolumn,pdflatex,showpacs,superscriptaddress]{revtex4}
\usepackage{amsmath, amsthm, amssymb,float}
\usepackage{amsfonts}
\usepackage{graphicx}
\usepackage{hyperref}
\usepackage{dcolumn}
\usepackage{bm}
\usepackage{textcomp}
\usepackage[normalem]{ulem}
\usepackage{mathrsfs}
\usepackage{color}

\newcommand{\tsp}{\hspace{3mm}}
\usepackage{epsfig}
\usepackage{textcomp}
\usepackage{epstopdf}

\begin{document}

\title{Laser ion acceleration from tailored solid targets with micron-scale channels}

\author{K. V. Lezhnin}
\email{klezhnin@princeton.edu}
\affiliation{Department of Astrophysical Sciences, Princeton University, Princeton, NJ 08544, USA}
\author{S. V. Bulanov}
\affiliation{Institute of Physics of the ASCR, ELI-Beamlines, Na Slovance 2, 18221
Prague, Czech Republic}
\affiliation{Kansai Photon Science Institute, National Institutes for Quantum
and Radiological Science and Technology, 8-1-7 Umemidai, Kizugawa-shi, Kyoto 619-0215, Japan}

\date{\today}

\begin{abstract}
Laser ion acceleration is a promising concept for generation of fast ions using a compact laser-solid interaction setup. In this study, we {theoretically} investigate the feasibility of ion acceleration from the interaction of petawatt-scale laser pulses with a structured target that embodies a micron-scale channel filled with relativistically transparent plasma. Using 2D and 3D Particle-In-Cell (PIC) simulations and theoretical estimates, we show that it is possible to generate GeV protons with high volumetric charge and quasi-monoenergetic feature in the energy spectrum. We interpret the acceleration mechanism as a combination of Target Normal Sheath Acceleration and Radiation Pressure Acceleration. Optimal parameters of the target are formulated theoretically and verified using 2D PIC simulations. 3D PIC simulations and realistic preplasma profile runs with 2D PIC show the feasibility of the presented laser ion acceleration scheme for the experimental implementation at the currently available petawatt laser facilities.
\end{abstract}

\maketitle

\section{Introduction}
While modern laser facilities { have a potential of reaching} ultra-high intensities up to $10^{24} ~\rm W/cm^2$ \cite{Danson2019}, delivering laser fields up to $1\rm GV/\mu m$, acceleration of charged particles using laser-target interaction becomes more of an interest. Highly energized charged particle beams have a broad range of applicability \cite{Daido2012}: in imaging \cite{Mackinnon2004}, medicine \cite{BK2002,Bulanov2002}, {controlled nuclear fusion \cite{MROTH}, and nuclear physics \cite{MNISHI}}. Beams of charged particles may reach ultrarelativistic energies, with the current record of electron bunches being accelerated up to $\sim 10$ GeV \cite{Gonsalves2019} using state-of-the-art Laser Wake Field Acceleration (LWFA) mechanism \cite{LWFA}. Ion acceleration is also estimated to be efficient from theory and simulations \cite{Esirkepov2006,Macchi2013,SVB2014,SSB2016}, but the experimental research reports the saturation of the maximum attainable ion energies on 100 MeV level \cite{Higginson2018}. Up and coming lasers with the peak powers reaching $10\, \rm PW$ \cite{ELINP,ELIBL,APOLLON} may help to overcome this level of ion energies, but the need for the theoretical understanding of possible limiting factors still exist. Therefore, a more detailed theoretical understanding of laser ion acceleration schemes, incorporating such physics as prepulse effects \cite{Kaluza2004,Esirkepov2014, PROHAD2020a}, field ionization \cite{MNISHI20}, oblique incidence \cite{Ferri2020}, pointing stability \cite{Gray2001}, and radiation reaction effects \cite{MTAMB2010}, is necessary for successful experimental delivery of high energy ion beams on a new generation of petawatt laser facilities.

On the theory side of laser ion acceleration, there are a few major mechanisms being discussed recently. The current state-of-the-art mechanism is Target Normal Sheath Acceleration (TNSA) (see \cite{SWILKS}, review articles \cite{Daido2012, Passoni2010, Macchi2013, SVB2014}, and references therein), which is realized by a build-up of an electrostatic field on the rear side of the thick target due to abundance of hot electrons generated by laser interaction with the front of the target. Accelerating electric field is known to be proportional to $\propto (T_{\rm e,nth} n_{\rm e,nth})^{1/2}$, with $T_{\rm e,nth}$ and $n_{\rm e,nth}$ denoting hot electron temperature and density, respectively, and multiple efforts are made in order to increase both hot electron population properties \cite{Liu2012,YOGO2016, YOGO2017, Zou2017,Zou2019}. A very promising maximum ion energy scaling with laser pulse power is provided by Radiation Pressure Acceleration (RPA) \cite{Esirkepov2004,SVBulanov2010}, which was observed experimentally \cite{SKAR2008, SKAR2012, Henig2009}. Multiple other mechanisms are also discussed, such as Coulomb Explosion \cite{Fourkal2005}, Magnetic Vortex Acceleration (MVA) \cite{Kuznetsov2001,FUKUDA2009,SSBulanov2010,Park2019}, Shock Acceleration \cite{Fiuza2012}, and combinations of these \cite{SSBulanov2008}.

Recently, solid-state targets started to gain more interest for electron acceleration \cite{Snyder2019,Wang2020a}, ion acceleration \cite{Liu2012,Zou2017,Zou2019} and radiation sources, such as X-ray \cite{Rousse1994,Andriyash2014} and $\gamma$-ray generation \cite{Nakamura2012,Ridgers2012}. In principle, higher density targets may lead to higher densities of fast electrons \cite{Zou2017,Zou2019} and better retention of fast electrons around laser-solid interaction spot \cite{Kluge2010}, which should benefit such acceleration schemes as TNSA and MVA. On the other hand, solid densities are generally opaque for optical laser pulses, which suppresses laser absorption.

This is where structured solid targets come into play. Structured targets may provide better laser-target coupling \cite{DM2012,Bailly2020}, edge field amplification \cite{Askaryan1983}, laser guidance \cite{Wang2014,SSBulanov2015}, and self-consistent ion injection into acceleration scheme \cite{MURAKAMI}. For instance, in \cite{Liu2012}, a solid target with holed conical opening and concave rear side with a proton layer doping was considered. The acceleration mechanism was attributed to a combination of TNSA and additional acceleration by the electric field of focused protons. Conical opening enhanced TNSA by a more effective hot electron generation on the rear side. A similar target, but with a plane rear side and comprised of high-Z ions was also discussed in \cite{Zou2017}. High-Z ions and microchannel structure were implemented to improve hot electron generation and avoid laser filamentation, respectively. We note that using thin foil targets with holes for laser ion acceleration has been actively studied theoretically and experimentally in Refs. \cite{PSIKAL2016, PROHAD2020b, CANT2021}. Microchannel target filled with relativistically transparent foam was considered in \cite{Arefiev2018}. Laser pulse was tightly focused into the channel, propagated through relativistically transparent plasma while delivering significant energy to electrons from the channel filling and solid target walls, and exited from the rear side. The fastest ions were generated on the rear side of the target at the moment when defocusing laser pulse started to exit the channel. Channel target filled with relativistically transparent foam was also considered in \cite{Stark2016,Jansen2018,He2021,Rinderknecht2021} for efficient generation of $\gamma$ rays via synchrotron emission of fast electrons in quasi-static MegaTesla-scale magnetic field generated by laser-foam interaction. These targets are experimentally available and provide flexibility for the particular experimental needs \cite{Snyder2019,Bailly2020,Li2021}.

In this paper, we explore laser ion acceleration from structured solid targets filled with relativistically transparent plasma by means of Particle-In-Cell (PIC) simulations and theoretical estimates. We find optimal conditions for high energy proton generation theoretically and verify them using comprehensive 2D PIC scans. The acceleration mechanism is interpreted as a combination of TNSA and RPA. We also conduct 3D PIC simulations and address such experimentally relevant questions as prepulse physics, oblique incidence, pointing stability, and the role of field ionization of the target. The role of the channel and solid target electrons is also discussed, as well as the role of radiation reaction (RR) for higher laser pulse powers. Finally, we discuss the scalability of the discussed acceleration scheme to the parameters of currently available channel targets \cite{Snyder2019,Bailly2020,Rinderknecht2021}.

This paper is structured as follows. Section II focuses on theoretical estimates for the maximum energy of protons from classical electrodynamics with and without the inclusion of radiation reaction force. Optimal target conditions for the given laser pulse parameters are also derived. Section III is devoted to the discussion of the simulation setup. In Section IV, we discuss the results of our 2D and 3D PIC scans and compare them with our theory. Other important aspects, such as prepulse effects, field ionization, oblique incidence, and pointing stability, are also addressed.
Finally, we conclude with Section V by discussing our main results and comparing them with the literature.

\section{Theory of ion acceleration from channel targets}

First, let us recall the concept of relativistic transparency, which is important for the considered ion acceleration scheme. As is well known, in a non-relativistic case, the electromagnetic wave does not propagate into cold unmagnetized plasma if $\omega_{\rm pe}\geq \omega_0$. 
Here, $\omega_{\rm pe}^2=4\pi n_ee^2/m_e$ is the {square of the } plasma frequency for a non-relativistic case. Thus, since in our case the channel density is $\sim10^{0}-10^{1} n_{\rm cr}$, where $n_{\rm cr} = m_e \omega_0^2/4 \pi e^2$ is the critical density for the electromagnetic wave of frequency $\omega_0$, we should have expected from non-relativistic considerations that the laser should reflect from the target. However, the relativistic motion of electrons relaxes the wave penetration condition due to an additional factor in the denominator of plasma frequency: 
$\omega_{\rm pe,rel}^2 = 4 \pi n_ee^2/\langle \gamma_e \rangle m_e=\omega_{\rm pe}^2/\langle \gamma_e \rangle$, with $\langle \gamma_e \rangle$ being an average electron gamma factor, that effectively decreases the threshold for laser pulse propagation in classicaly overcritical plasma. Since electrons in the laser pulse gain average energy of $m_ec^2 a_0$, 
where $a_0=eE_0/m_e\omega_0c$ is the {normalized} amplitude of the laser pulse, we expect $\langle \gamma_e \rangle \approx a_0$, leading to $\omega_{\rm pe,rel} = \omega_{\rm pe}/\sqrt{a_0}$ (see Ref. \cite{AKHPOL}). For tightly focused petawatt-scale laser pulses, it will lead to the relativistic transparency of the channel for laser pulse, thus providing efficient conditions for laser-target coupling.

\begin{figure}
    \centering
    \includegraphics[width=\linewidth]{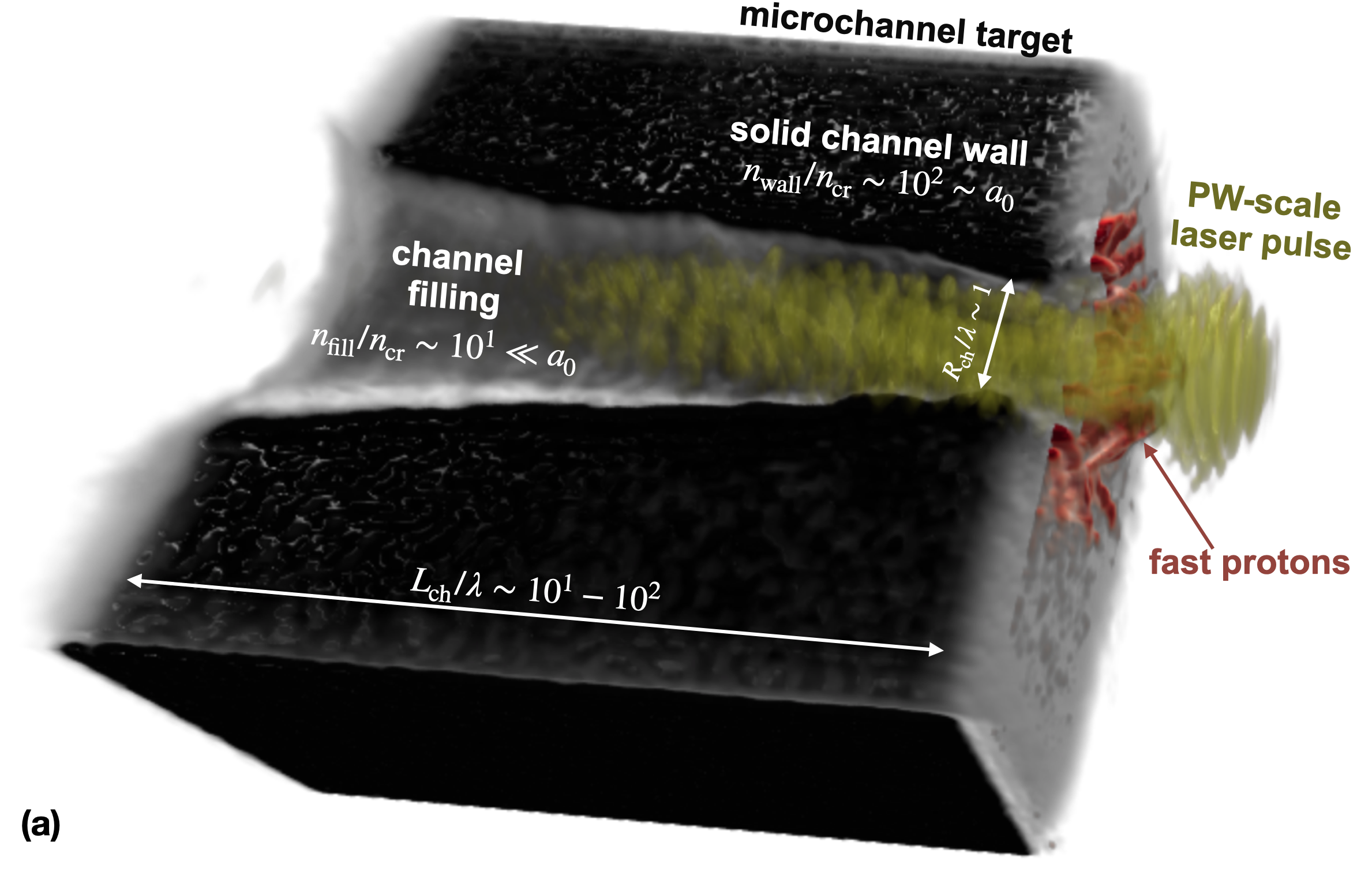} \\
    \includegraphics[width=\linewidth]{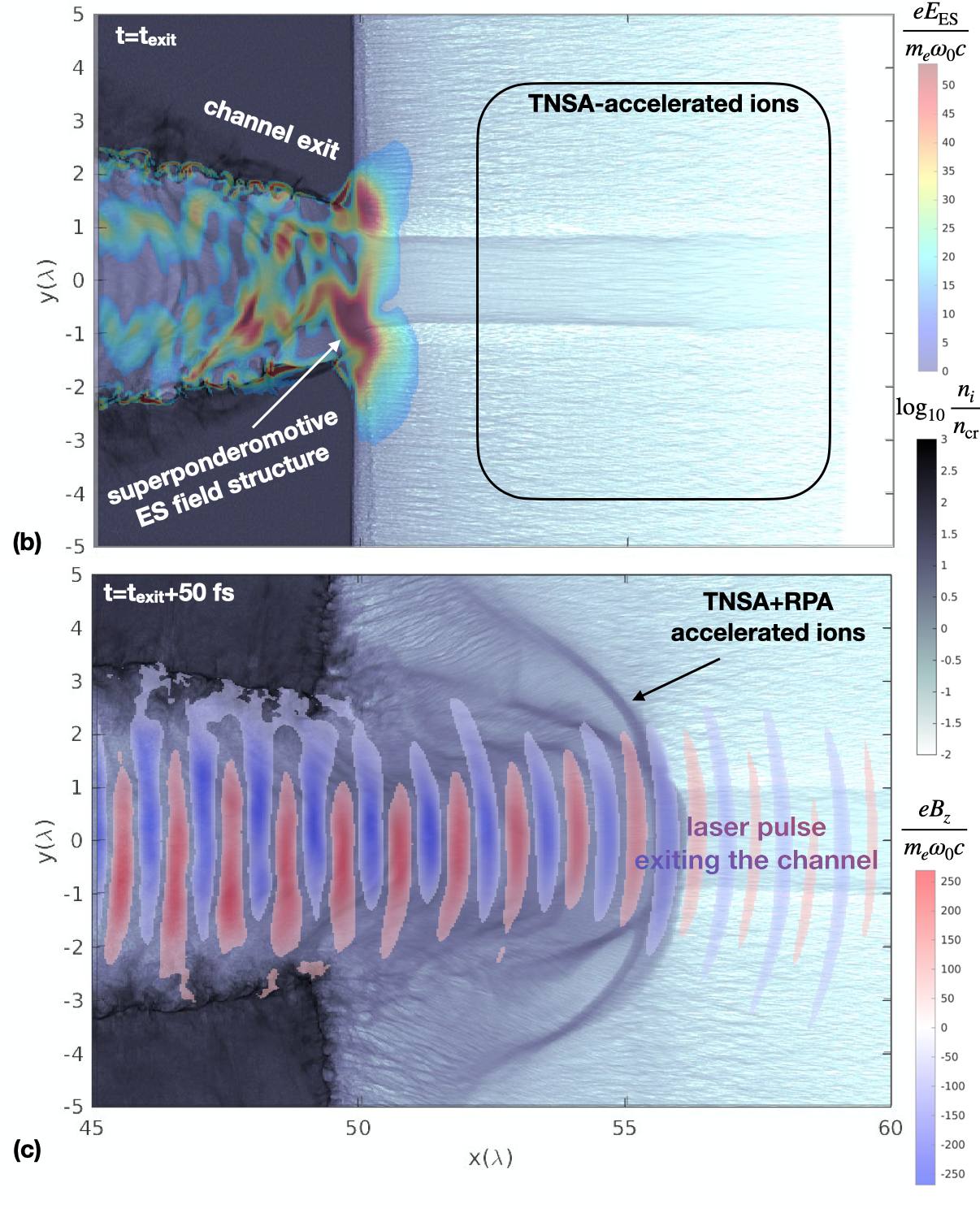}
    \caption{(a) Result of 3D PIC simulation with $P=10$ PW, $L_{\rm ch}=40 \mu \rm m$, and $n_{\rm ch}=10 n_{\rm cr}$. Primary free parameters of the problem are indicated. (b),(c) Illustration of acceleration scheme from 2D PIC for $P=10$ PW and $L_{\rm ch}=40 \mu \rm m$: electrostatic field, evolution of ion density, and laser field at time of laser pulse exiting the channel ($t=t_{\rm exit}$) and 50 femtoseconds later.
}
    \label{fig:scheme}
\end{figure}

Following \cite{SSBulanov2010}, let derive the optimal condition for ion acceleration. First, the total energy of laser pulse is:

\begin{equation}
    \mathcal{E}_{\rm las,0} = I_0 a_0^2 \pi w^2 \tau_{\rm las},
    \label{eqn:Elas}
\end{equation}

\noindent where $I_0 = 1.384\cdot 10^{18}\, \rm W/cm^2$ and $\tau_{\rm las}$ is the duration of the laser pulse, and $w$ is the waist of the laser pulse. This energy is fixed for a particular laser pulse. At the same time, total energy of the electrons in the channel after the partial absorption of the pulse may be estimated by the formula:

\begin{equation}
    \mathcal{E}_{\rm ele} = m_e c^2 a_0 n_{\rm ch} \pi R_{\rm ch}^2 L_{\rm ch},
    \label{eqn:Eele}
\end{equation}

\noindent where $n_{\rm ch}$, $R_{\rm ch}$, and $L_{\rm ch}$ are channel electron number density, radius, and length, respectively. By equating Eqns.~\ref{eqn:Elas} and \ref{eqn:Eele}, we get an optimal condition for ion acceleration in the case of magnetically assisted TNSA (MVA) \citep{SSBulanov2010,Park2019}. The leftover energy of the pulse after propagating in the channel (if any) will be responsible for the radiation pressure acceleration. Let's assume that the energy dissipation happens in such a way that only affects the field amplitude. The energy of the laser pulse after exiting the channel will look like the following:

\begin{equation}
    \mathcal{E}_{\rm las,ch} = \mathcal{E}_{\rm las,0}-\mathcal{E}_{\rm ele}=I_0 a_{\rm ch}^2 \pi w^2 \tau_{\rm las}.
    \label{eqn:Elasch}
\end{equation}

\noindent In the case of high laser intensities, we may also want to include energy losses due to radiation reaction in consideration. As is well known {\cite{LLAD}, in the near-critical density plasma} for dimensionless field amplitudes 
{
\begin{equation}
a_0 \geq \left(\frac{3\lambda m_e c^2}{4\pi e^2}\right)^{1/3}, 
\label{eqn:a0RR}
\end{equation}
(for $\lambda = 1\mu$m wavelength laser the radiation intensity should be above $ 10^{23}\, $W/cm$^2$)
} the radiation reaction force becomes important. The energy lost by electromagnetic pulse propagating in plasma may be estimated as follows. A single electron maximum radiation power may be calculated as $\mathcal{P}_{\rm RR} = eEc$. Total energy lost to radiation would be:

\begin{equation}
    \mathcal{E}_{\rm RR}= \mathcal{P}_{\rm RR} n_{
    \rm ch} \pi w^2 \tau_{\rm las} L_{\rm ch} = \mathcal{E}_{\rm ele}.
    \label{eqn:EnRR}
\end{equation}

\noindent The remaining laser pulse energy that will accelerate ions via radiation pressure:

\begin{equation}
    \mathcal{E}_{\rm las,ch+RR} = \mathcal{E}_{\rm las,0}-\mathcal{E}_{\rm ele}-\mathcal{E}_{\rm RR}=I_0 a_{\rm ch+RR}^2 \pi w^2 \tau_{\rm las}.
    \label{eqn:ElaschRR}
\end{equation}

Now, let's optimize the maximum ion energy obtainable in the considered ion acceleration scheme. For simplicity, we ignore the RR losses for now.
The maximum energy gained from TNSA-like acceleration may be estimated as (following \cite{SWILKS})

\begin{equation}
\Delta \mathcal{E}_{\rm TNSA}\approx \alpha T_{\rm e,nth} = \alpha m_ec^2( \sqrt{1+a_0^2}-1)\approx \alpha m_ec^2 a_0,
\label{eqn:EnTNSA}
\end{equation}

\noindent 
where $\alpha$ is a dimensionless constant larger than 1 which accounts for the finite size of the accelerating field, 
energy cutoff of non-thermal electron energies in the rear side of the channel, 
and possible superpondermotive electron temperatures (e.g., see \cite{Zou2017}). 
This constant is to be determined by simulations. 
{ In the limit of the ultrarelativistic ion energy, the energy gained from RPA can be estimated 
as (e.g. see Ref. \cite{SVB2014})

\begin{equation}
\Delta \mathcal{E}_{\rm RPA} \approx m_e c^2 a_0^2 \frac{n_{\rm cr} c \tau_{\rm las}}{n_{0} l_0},
\label{eqn:EnRPA}
\end{equation}
where $n_0$ and $l_0$ are density and thickness of the foil target, and $n_{\rm cr}=m_e \omega_0^2/4\pi e^2$}. 
In our case, there is no foil target, but the hole boring by the laser pulse creates a dense foil-like structure with the thickness $\sim \lambda$ at the end of the channel. Assuming that this structure is comprised of the channel electrons, we write a condition on $n_0$ and $l_0$: $n_0 l_0 \approx L_{\rm ch} n_{\rm ch}$. In order to estimate the energy gain by an ion from RPA in a non-optimized target case, we multiply this expression by a function $\Xi=f(a_{\rm ch}, \frac{n_{\rm ch}}{n_{\rm cr}},\frac{L_{\rm ch}}{\lambda})$ that will isolate an optimal regime of RPA ion acceleration, $a_0 \approx n_{\rm ch}/n_{\rm cr} L_{\rm ch}/\lambda$, and exponentially damp the acceleration away from "the resonance condition", i.e. it is equal to one for optimal RPA condition and quickly tends to zero outside of it. Finally, we multiply this expression by dimensionless constant $K$, which controls the efficiency of RPA acceleration (i.e., the reflectivity of the foil accelerated by RPA) and will be determined from simulations as well. The total energy gain by a single proton may be estimated as:
{
\begin{equation}
\begin{split}
   & \mathcal{E}_{\rm max} = \Delta \mathcal{E}_{\rm TNSA}+\Delta \mathcal{E}_{\rm RPA}  \approx \\  
    & m_ec^2\left\{
    \alpha a_0 +Ka_{\rm ch}^2 \frac{n_{\rm cr}}{n_{\rm ch}}\frac{c\tau_{\rm las}}{L_{\rm ch}}\Xi \left(a_{\rm ch}, \frac{n_{\rm ch}}{n_{\rm cr}},\frac{L_{\rm ch}}{\lambda}\right) \right\},  
\end{split}
\label{eqn:maxenergy}
\end{equation}
}
\noindent where

\begin{equation}
    a_{\rm ch}^2 = a_0^2 - a_0\frac{m_ec^3 n_{\rm cr}}{I_0}\frac{R_{\rm ch}^2}{w^2}\frac{n_{\rm ch}}{n_{\rm cr}}\frac{L_{\rm ch}}{c\tau_{\rm las}}
\end{equation}

\noindent is the laser field dimensionless amplitude after the depleted laser pulse exits the channel (here we neglect RR energy losses). Maximizing $\mathcal{E}_{\rm max}$ will give an optimal condition for ion acceleration. Since the energy gain is dominated by the RPA mechanism, we may approximately claim that the optimal condition is

\begin{equation}
     a_{\rm ch}\approx\frac{n_{\rm ch}}{n_{\rm cr}}\frac{L_{\rm ch}}{\lambda} .
    \label{eqn:optcond}
\end{equation}
Now, let us describe a simple model to explain the maximum ion energy scaling with time. As noted above, we assume that the ions are first accelerated by TNSA electric field, and further accelerated by RPA, and the total energy gain is $\mathcal{E}_{\rm max} =\Delta \mathcal{E}_{\rm TNSA}+\Delta\mathcal{E}_{\rm RPA}$. We assume TNSA acceleration to be instantaneous, and the evolution of ion energy under the radiation pressure is calculated as in Refs. \cite{Esirkepov2004,SVBulanov2010}: we start with a 1D model and write down the equation of motion of plasma under the influence of radiation pressure:

\begin{eqnarray}
    \partial_\tau p = \mathcal{P} \frac{1-\beta}{1+\beta},     \label{eqn:RPA1Da} \\
    \beta = \frac{p}{\sqrt{1+p^2}}.
    \label{eqn:RPA1Db}
\end{eqnarray}

\noindent Here, $p$ is normalized to $m_ic$, $\tau \equiv t/T_0$, and the radiation pressure is given by 
\begin{equation}
\mathcal{P} = K\frac{m_e}{m_i}a_0^2\frac{n_{\rm cr}}{n_e}\frac{\lambda}{l_0}.
\end{equation}
 $K$ here is a free dimensionless parameter (same as in Eqn.~\ref{eqn:maxenergy}) that controls the laser-target interaction efficiency. Asymptotic solution for ultrarelativistic ions, $p \propto t^{1/3}$, is well known \cite{Esirkepov2004}. However, here we do not expect ions to gain ultrarelativistic energies, so we solve Eqns.~\ref{eqn:RPA1Da}-\ref{eqn:RPA1Db} with $p(t=0)=p_{\rm TNSA} = \sqrt{2 m_i \mathcal{E}_{\rm TNSA}}$, i.e. the initial ion momentum is gained from TNSA fields. We see that the channel parameters and laser amplitude appear in $\mathcal{P}$, and $p(0)$ also implicitly depends on channel parameters. We fit the resulting $p(t;\alpha,K,t_0)$ trajectories from simulations with free parameters $\alpha,K,t_0$. It turns out that the model describes $p(t)$ from simulations fairly well, with $\alpha \sim 3-5$, $K \sim 0.1-0.3$, $t_0 \approx t_{\rm exit}$, with $t_{\rm exit}$ being the time of laser pulse exiting the channel from the rear end.

\section{Simulation setup}

\begin{figure}
    \centering
    \includegraphics[width=\linewidth]{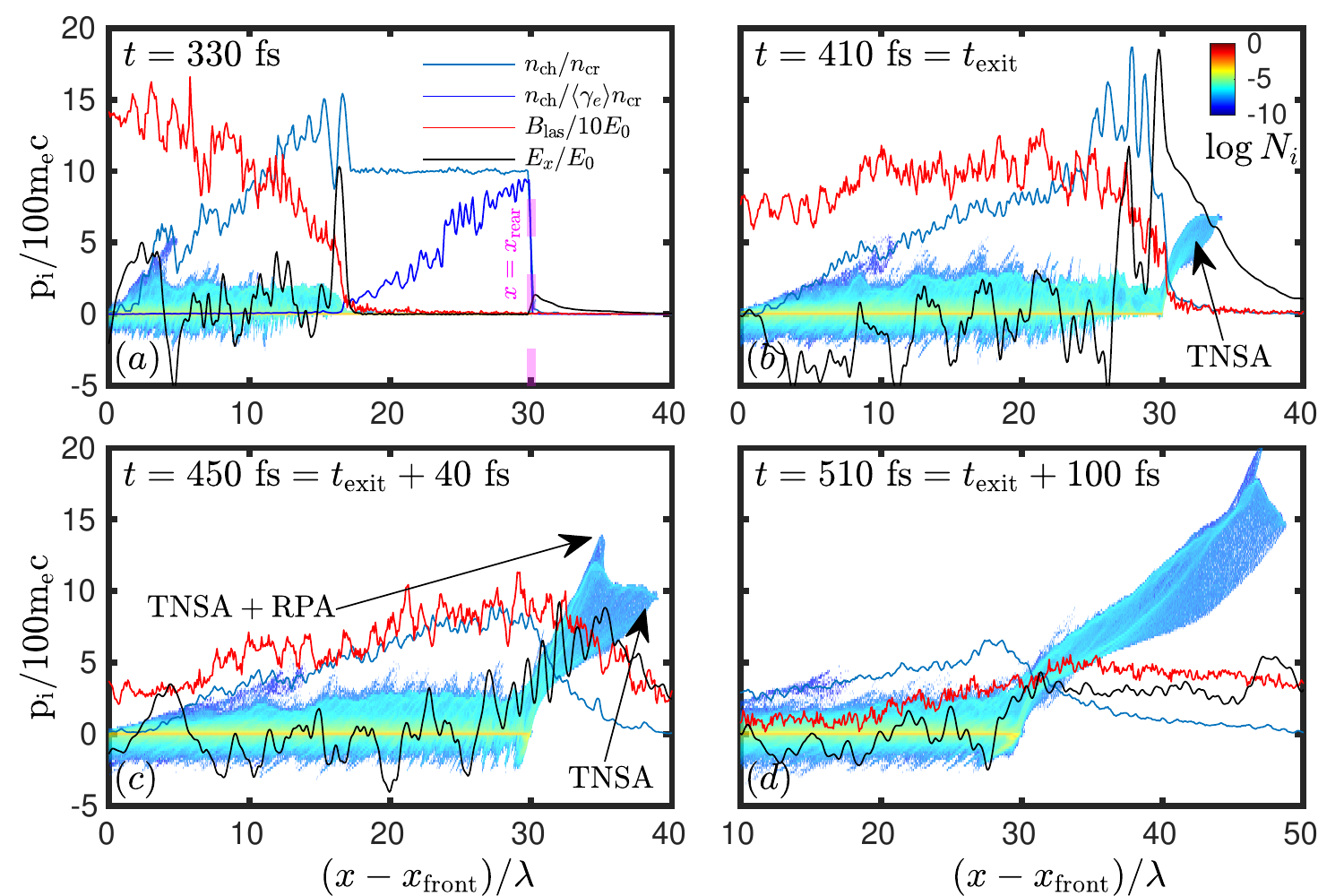}
    \caption{Evolution of ion phase space in $x-p_x$ coordinates (colormap), along with 1D profiles of electron density (light blue), electron density normalized to average electron gamma (dark blue), laser envelope (red), and longitudinal electric field (black). (a) t=330 fs, laser pulse propagates in the channel, a significant laser-electron coupling is seen - electrons are not evacuated from the channel - the channel is relativistically transparent - sustaining a significant electron density well above $n_{\rm cr}$; (b) t=410 fs, laser pulse reaches the rear end of the target, accelerating electric field builds up on the rear part of the target, TNSA accelerated ions are seen; (c) t=450 fs, laser pulse leaves the channel and provides RPA acceleration of ions; rapid acceleration of ion filament from the rear side of the channel is seen (annotated as TNSA+RPA); (d) t=510 fs, the most rapid phase of ion acceleration is over (see Figure 4a, green line), but ions continue to gradually gain more energy via RPA.}
    \label{fig:phspace} 
\end{figure}

\begin{figure}
    \centering
    \includegraphics[width=\linewidth]{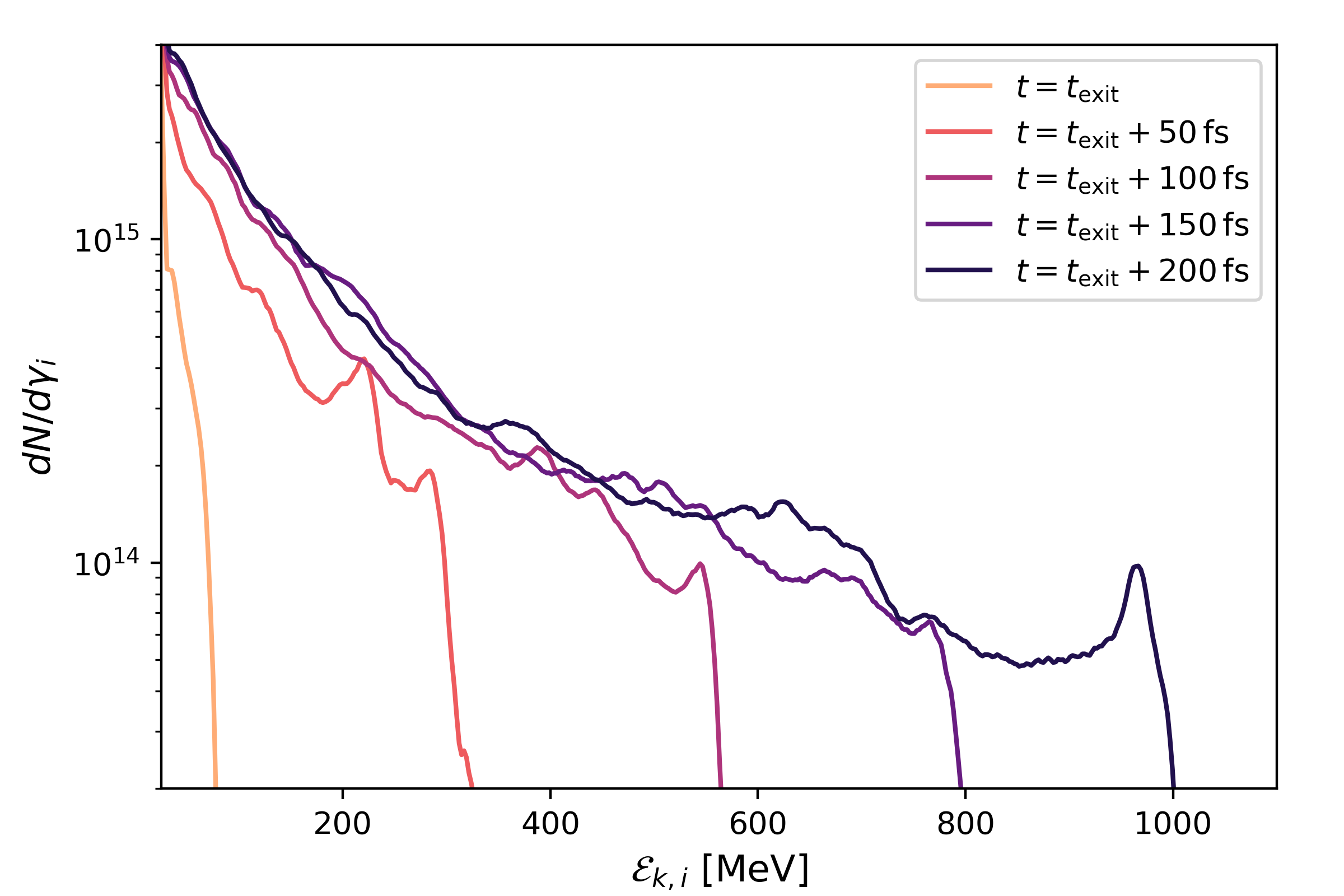}
    \caption{Proton spectrum evolution for the simualtion with $P=1~ \rm PW$, $L_{\rm ch}=50~ \mu \rm m$, $n_{\rm ch}=10 n_{\rm cr}$. The development of the high-energy spectrum and monoenergetic-like peak is seen after the laser pulse starts to exit the channel.}
    \label{fig:spec}
\end{figure}

To check the theoretical considerations from the previous Section and consider a more realistic physical scenario, 
we perform 2D and 3D particle-in-cell (PIC) simulations using the code EPOCH \cite{EPOCH1, EPOCH2}. 
Numerical setup and illustration of typical 2D/3D simulation result is shown in Figure 1. All the parameters we scan on are summarized in Table \ref{Table2Dscan}. For 2D runs, we consider Gaussian laser pulses with laser wavelength $\lambda=1\mu \rm m$, laser durations $\tau=30 ~\&~ 150$ fs, $w=1.1 - 15 \, \mu \rm m$ waist, and linear polarization ($B_z$ is out of simulation plane $x\text{-} y$) focused onto the channel entrance 
at $x=10 \mu \rm m$. The laser pulse power spans from 0.3 to 30 PW, covering the range of dimensionless amplitudes, $a_0$, from $25$ to $850$ (peak intensities range from $10^{21}$ to $3 \cdot 10^{23}$ W/cm$^2$). The target locates between $x=10\,\mu \rm m$ and $10 \mu \rm m$+$L_{\rm ch}$, where $L_{\rm ch}$ is the channel length, which varies from 10 to 100 micron. Solid wall density equals to $100-300 n_{\rm cr}$. The channel has the radius $R_{\rm ch}=1-5 \mu \rm m$, and is filled with uniform plasma with $n_{\rm ch} = 0-40 n_{\rm cr}$. The plasma is comprised of electrons and protons with zero initial temperature. The simulation box dimension is $(160 \lambda + L_{\rm ch}) \times 30 \lambda$ with the numerical resolution of 60 grid nodes per $\lambda$. The resolution ensures that typical plasma wavelength, $\lambda_{\rm pe}=2\pi c/\omega_{\rm pe}$, is resolved with 6 grid nodes. The boundary conditions are outflow for both axes. The number of particles per cell is 20-80 per species. We conduct runs with radiation reaction (RR) terms turned on and off to see its influence on ion acceleration.

\begin{table}
\centering
\caption{2D PIC scan parameters}
\begin{tabular}{lrr}
\tableline\tableline
& range \\

Laser parameters: \\
\tsp Peak power, $P$, PW & $0.3-30$ \\
\tsp Pulse duration, $\tau_{\rm las}$, fs & 30,\,150 \\
\tsp Waist, $w$, $\mu $m & $1.1-15$ \\
\tsp Laser wavelength, $\lambda$, $\mu $m & $1$ \\
\tsp Contrast, $I_{\rm prepulse}/I_{\rm max}$ & $0.0, ~10^{-6}-10^{-3}$
\\
\tsp Prepulse duration, ps & 1 \\
\\
Target parameters: \\
\tsp Channel radius, $R_{\rm ch}/\lambda$ &  $1-10$\\
\tsp Channel length, $L_{\rm ch}/\lambda$ & $10-100$\\
\tsp Filling density, $n_{\rm ch}/n_{\rm cr}$ & 0-40 \\
\tsp Solid wall density, $n_{\rm wall}/n_{\rm cr}$ & 100,300 \\
\tsp Target front cut angle, $^\circ$ & 0,15,45 \\
\\
General parameters: \\
\tsp Simulation box size, $\lambda \times \lambda$ & $200 \times 30$ \\
\tsp Grid resolution, 1/$\lambda$ & 60 \\
\tsp Particle resolution, ppc & 20,40,80\\
\tsp Total simulation time, ps & 1.5-2.5 \\
\tsp Radiation reaction term & on \& off \\
\tsp Field ionization & on \& off \\

\end{tabular}
\label{Table2Dscan}
\end{table}

To address the case with realistic target material, e.g., solid Kapton substrate and CH foam as the channel filling \cite{Rinderknecht2021}, we considered CH targets with $L_{\rm ch}=20-50 \lambda$, $R_{\rm ch}=1.8 ~\lambda$, $w=2.2 \lambda$, $n_{\rm e, wall}=300 n_{\rm cr}$, $n_{\rm e,ch}=10-30 n_{\rm cr}$, and fully ionized C and H atoms.

We also considered oblique incidence by adding a cut to the front side of the target. Oblique incidence ensures the absence of the backreflection of the laser pulse, which is safer for possible application on laser facilities \cite{Snyder2019,Bailly2020}. We consider a cut with $10^\circ$ and $45^\circ$ angles on the front of the target while keeping all other simulation parameters the same as described above.

For 3D simulations, following \cite{Arefiev2018}, we consider 1 PW and 10 PW, 150 fs Gaussian linear polarized pulses focused onto the channel target entrance onto the 2.2 micron spot. The considered target parameters are similar to ones in 2D simulations, with $L_{\rm ch}=20-30~ \mu \rm m$, $n_{\rm wall}/n_{\rm cr}=100$, $R_{\rm ch}=1.8 ~\mu \rm m$, and $n_{\rm ch}/n_{\rm cr}=10$, comprised of protons are electrons.

Finally, for auxiliary radiation hydrodynamics simulations using FLASH code, we inherited the LaserSlab simulation setup \cite{FLASH,FLASH1}, which considers the interaction of laser beam with the typical nanosecond laser pedestal parameters with the solid aluminum target. In our case, we conducted a set of analogous runs with only modifications being the modified density profile - we introduced a channel of $R_{\rm ch} =3~\mu \rm m$ at the axis of $R-z$ simulation plane in cylindrical coordinates - and considered a polystyrene (CH) target corresponding to $n_{\rm e, wall} = 300 n_{\rm cr}$  and $n_{\rm e, ch} = 20 n_{\rm cr}$, while also expanding the simulation box to $120~ \mu \rm m$ along $z$ axis, resulting in $40 \lambda \times 120 \lambda$ dimensions in $R-z$ space, with the channel located between $z=40 \lambda$ and $80\lambda$. The laser pulse has the wavelength of $1~\mu \rm m$, Gaussian transverse shape with the e-folding length of $3 ~\mu \rm m$, and focused onto the center of the channel entrance with normal incidence. The temporal profile of the laser pulse has a linear ramp of 0.1 ns from 0 to peak power and duration of 0.9 ns with the total simulation time being 1 ns. We varied the peak laser pedestal power, covering the range from $10^5$ to $10^{9}$ W. This corresponds to laser contrasts from $10^{-11}$ to $10^{-7}$ for 10 PW driver pulse. The resulting density snapshots from these simulations were mirrored around the $z$ axis, zoomed in to $-20 \lambda$ to $20 \lambda$ in the transverse direction, and inserted into 2D PIC code EPOCH to analyze the detrimental role of the prepulse in laser ion acceleration.

\section{Simulation results}

\begin{figure}
    \centering
    \includegraphics[width=\linewidth]{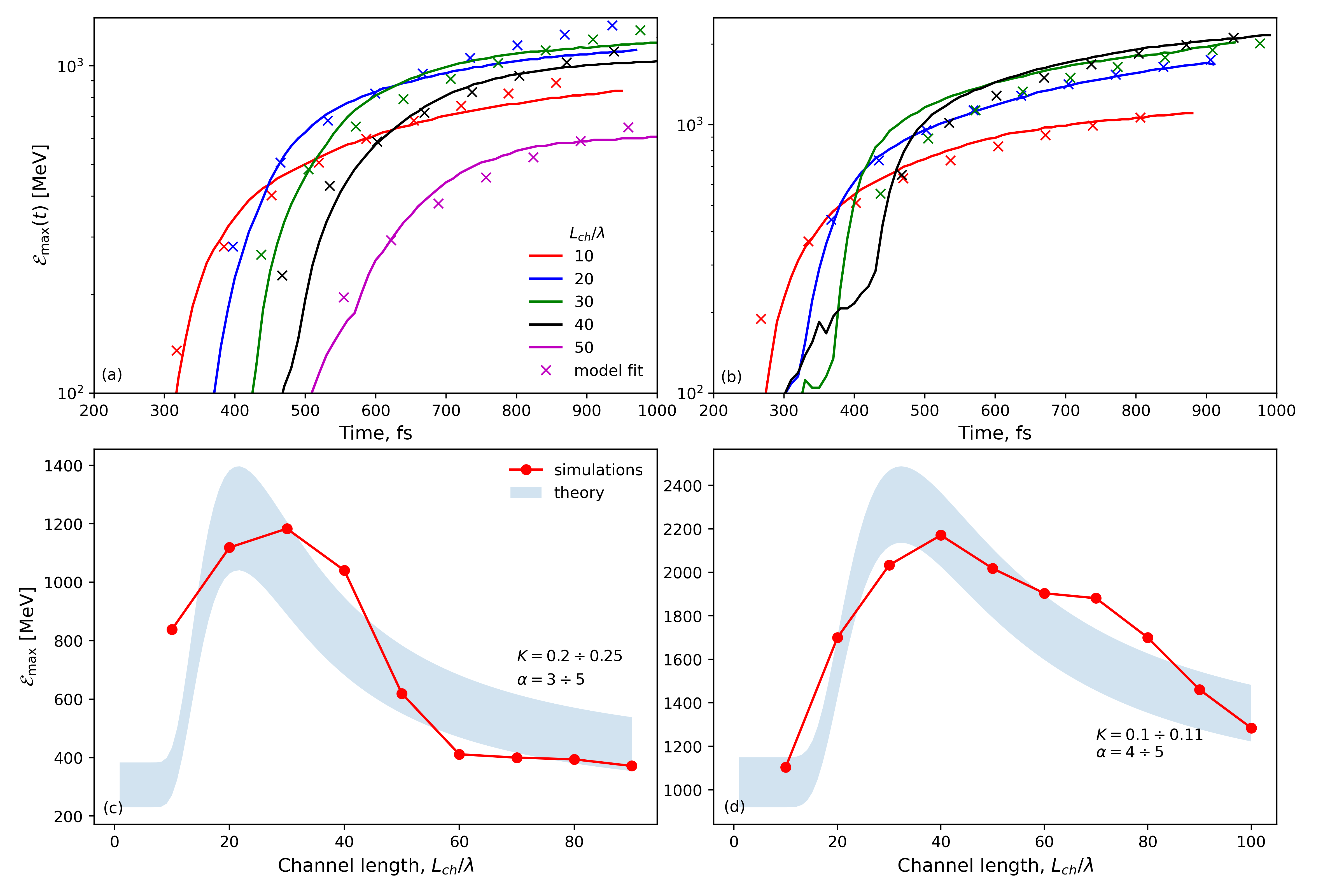}

    \caption{Time evolution of maximum ion energy (a and b) and channel length scan vs maximum ion energy (c and d) for 1 PW peak laser pulse power (left) and 10PW peak laser pulse power (right). Crosses and shaded region denote theoretical predictions for maximum ion energy temporal evolution and maximum ion energy as the function of channel length, respectively.}
    \label{fig:emax_vs_t_lch}
\end{figure}

\begin{figure}
    \centering
    \includegraphics[width=\linewidth]{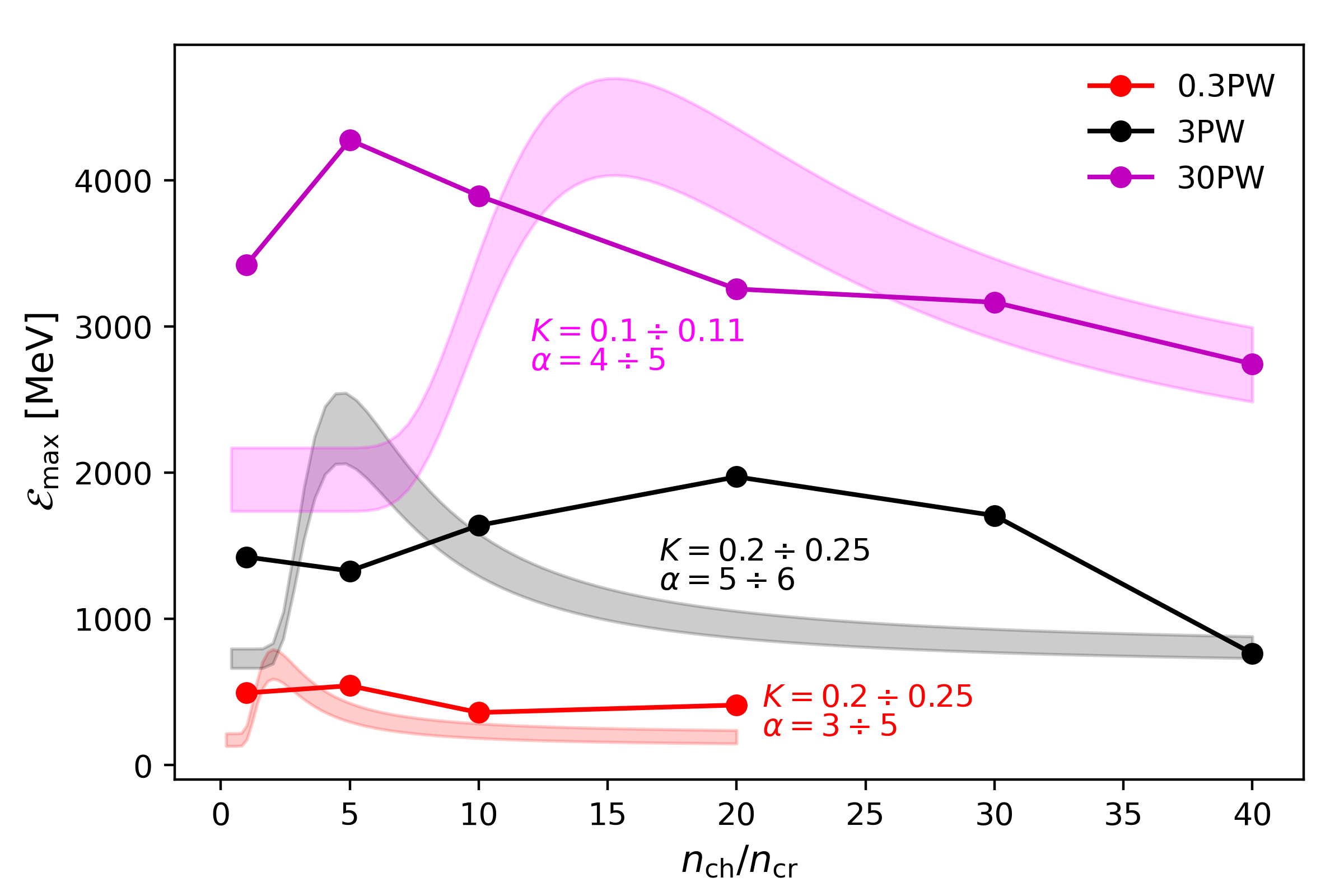}
    \caption{Dependence of maximum ion energy on channel filling density for pulses from 0.3 to 30 PW. Shaded regions illustrate theoretical predictions for maximum ion energies as the function of channel filling density.}
    \label{fig:emax_vs_nch}
\end{figure}

\begin{figure}
    \centering
    \includegraphics[width=\linewidth]{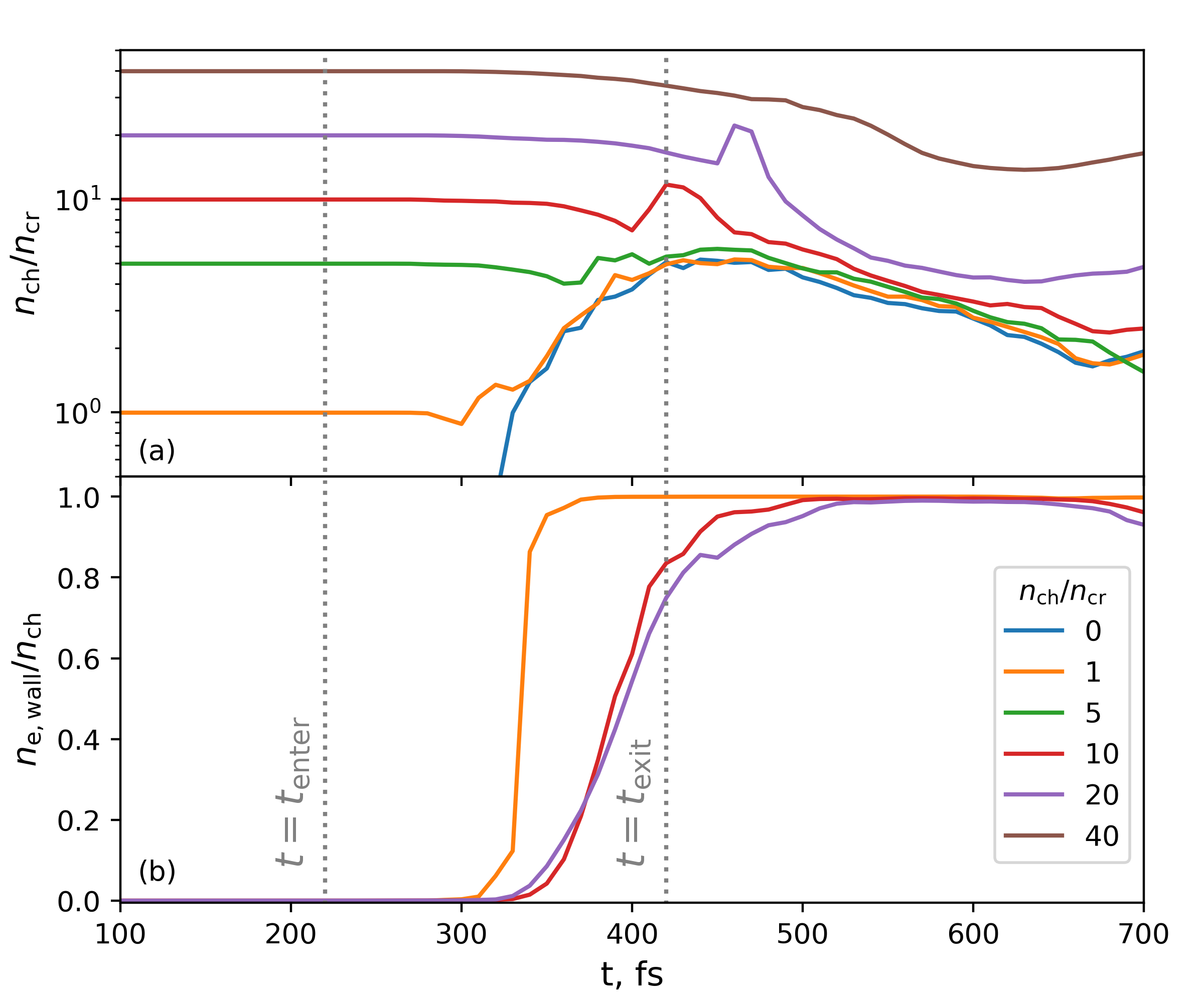}
    \caption{(a) Evolution of the average channel density at the exit of the channel for  $P = 1 \, \rm PW$ and (b) evolution of the fraction of the wall electrons in the average density at the exit of the channel for $P = 10 \, \rm PW$. The build-up of the universal density value for all simulations with $n_{\rm ch}/n_{\rm cr}\leq 10$ and dominance of the wall electrons contributing to ion acceleration are seen.}
    \label{fig:filling}
\end{figure}

\begin{figure}
    \centering
    \includegraphics[width=\linewidth]{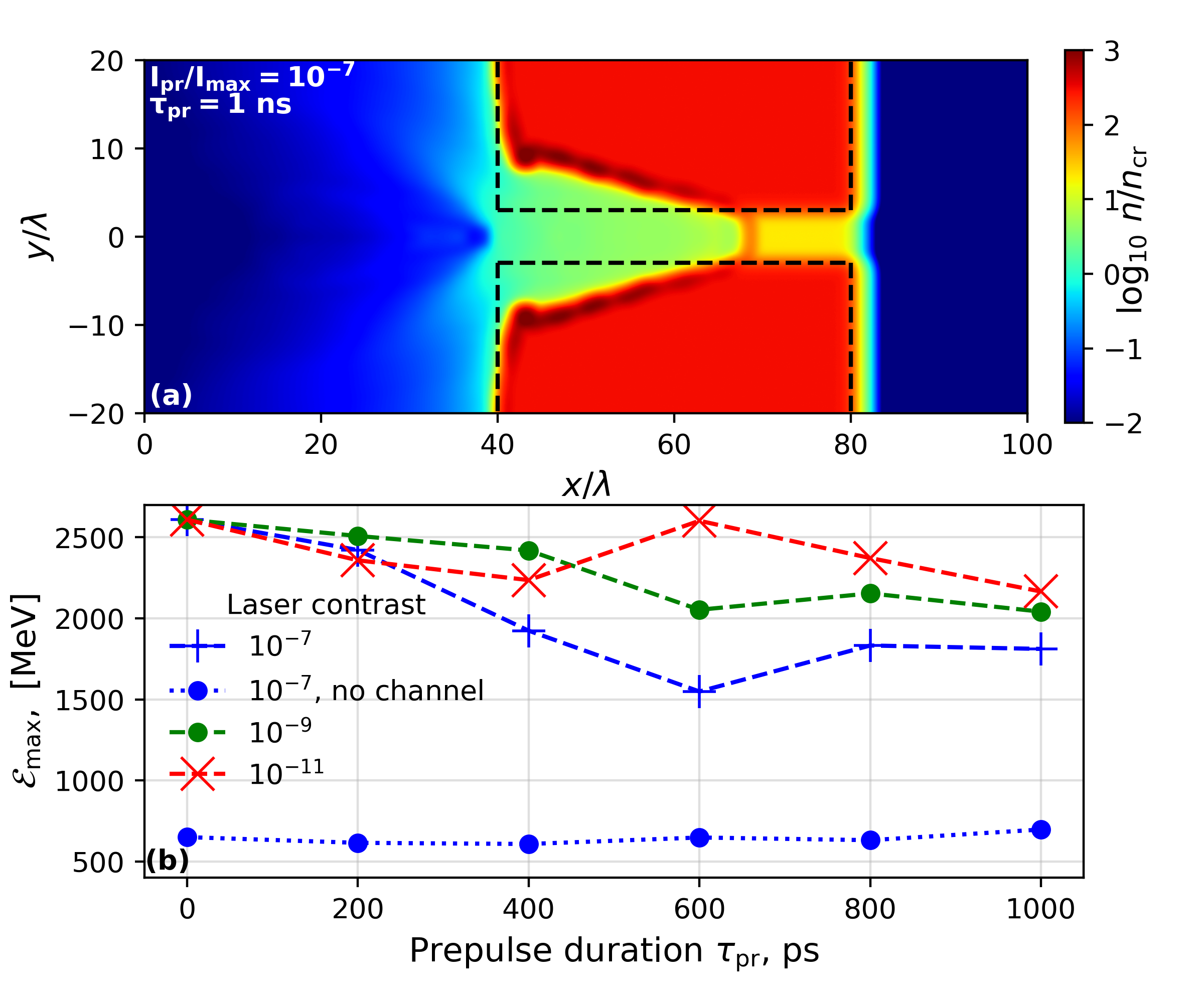}
    \caption{(a) Electron density snapshot from FLASH simulation of nanosecond pedestal-target interaction for the case of laser contrast of $10^{-7}$. Dashed black lines sketch the initial location of the channel structure. (b) Maximum ion energy dependence on pedestal duration for different laser contrasts ($10^{-11},~10^{-9},~10^{-7}$) and no channel case for laser contrast $10^{-7}$.}
    \label{fig:flash}
\end{figure}

First, let us discuss the typical 2D PIC simulation result for $P = 1$ PW, $n_{\rm ch}/n_{\rm cr}=10$, $L_{\rm ch}= 30 \,{\rm \mu m}$, $w=1.1~\mu \rm m$, and $R_{\rm ch} = 1 {\mu \rm m}$. Figure \ref{fig:phspace} illustrates the physics of the two-stage acceleration process. It combines 1D profiles of longitudinal electric field $E_x/E_0$ (averaged over $1 \mu m$ in the transverse direction, i.e. over the central half-channel; electric field is measured in $E_0 = m_e\omega_0c/e$), 1D envelope of the laser pulse at $y=0$, $B_z/10E_0$, 1D profile of electron density $n_e/n_{\rm cr}$ (averaged over $1 \mu m$ in the transverse direction as well) and $x-p_{\rm ix}$ phase space plot for t=330, 410, 450, 510 fs. The laser pulse is focused onto the front side of the channel, and since $n_e /\langle \gamma_e \rangle \ll n_{\rm cr}$ (dark blue line in Fig.2a), the pulse will propagate into the channel without significant backreflection. Also, since the channel density is relativistically transparent, it is not completely wiped out by ponderomotive forces and always stays well above non-relativistic critical density $n_{\rm cr}$, thus providing better laser-electron coupling (blue line on Figs.\ref{fig:phspace}a-d). Once the pulse reaches the rear part of the target at t=410 fs, we observe a build-up of a strong longitudinal electric field (predominantly, electrostatic) up to $E_x/E_0 \approx 20$. At this time, ion phase space exhibits an onset of TNSA-like acceleration at the rear end of the target (Fig. \ref{fig:phspace}b, $x \approx 45 \mu \rm m$; also Fig. 1b for a 2D density map of ions accelerated solely by TNSA). However, the fastest ions in the simulation are generated promptly at the time of laser pulse exiting the rear side of the channel (see Fig. \ref{fig:phspace}c, spike at $x\approx 45 \mu \rm m$). These ions are accelerated by TNSA field first, and then further accelerated by RPA (Fig.\ref{fig:phspace}d).

Figure \ref{fig:spec} represents the time evolution of proton spectrum for target with $L_{\rm ch}=50~ \mu \rm m$, $n_{\rm ch} = 10 n_{\rm cr}$ and laser pulse peak power of $P=1 \rm PW$. At the final time, $t=t_{\rm exit}+200 ~\rm fs$, there is a relatively flat spectrum in the high-energy range, with a peak around the maximum ion energy. Time evolution of the spectrum reveals that the peak in the ion spectrum is developed at the time of the laser pulse exiting from the rear side of the channel. Further acceleration is achieved by the direct acceleration of ions by laser pulse via RPA, as suggested by Fig.2d and Figs.4a,c.

Figure 4a,b shows maximum ion energy as a function of time for $1$ PW and $10$ PW laser pulses for the selected channel lengths from $10$ to $100~ \mu \rm m$. By fitting the curves from Fig.4a,b, we see that they are closely followed by the theoretical model described in Section II, and tend to have the same late-time scaling as RPA (measured scaling is $p_i \propto t^{0.37}$, which is close to RPA's $p_i \propto t^{1/3}$). Figures 4c,d demonstrate how maximum ion energy depends on $L_{\rm ch}$. It is seen that there is an optimal channel length, in accordance with Eqn.~\ref{eqn:optcond}. Our theoretical model fairly predicts an optimal channel length (shaded regions on Figs.~4c,d).

From Eqn.~\ref{eqn:optcond} it is also seen that there is an optimal channel density. Figure \ref{fig:emax_vs_nch} shows how maximum ion energy depends on channel filling density, $n_{\rm ch}$, for a fixed set of other parameters. While also predicting the existence of an optimal channel density, the agreement is worse than for a channel length, $L_{\rm ch}$. This may be explained by an effectively different channel density at the rear end of the channel, which is a combination of initial channel filling that stayed inside the channel and channel solid wall parts that were extracted by the intense laser pulse. Analyzing an average electron density at the rear end of the channel (i.e. inside the channel within $2 \lambda$ from the channel rear end), we found that at the time of the laser pulse exiting the channel, the electron density there turns out to be almost identical for initial channel density fillings in the range $n_{\rm ch}/n_{\rm cr} = 0-10$, see Figure \ref{fig:filling}. Simulations with tagged channel filling and wall electrons (Figure \ref{fig:filling}b) explicitly demonstrate that the contribution of channel filling is negligible in comparison to wall electrons that end up at the rear part of the channel. Additionally, the variance of the maximum ion energy with respect to channel filling density is less than $ 25 \%$ for $n_{\rm ch}\leq 30 n_{\rm cr}$. To conclude, the channel filling density plays a relatively minor role in ion acceleration, which implies relaxed requirements on the laser contrast for the described laser ion acceleration scheme.

To elaborate on the role of the laser contrast for the considered target, we perform an additional set of 2D PIC simulations, where a Gaussian prepulse of picosecond duration is added before the primary pulse, with the laser contrast, $I_{\rm prepulse}/I_{\rm max}$, varied from $10^{-6}$ to $10^{-3}$. While the duration of the prepulse may reach up to nanosecond duration \cite{Pathak2021}, which is beyond reach for the conventional PIC codes, significant damage to the target may be done by spontaneous pre-pulses of shorter duration, such as a considered picosecond prepulse \cite{Pathak2021}. We examine the cases of $P=1 \& 10$ PW, $n_{\rm ch}/n_{\rm cr}=0,1,10,20$. The variations in maximum ion energy with contrast are no more than 25\%, with higher contrast runs typically overperforming the corresponding runs with lower contrast. The overall acceleration mechanism seems to be unaffected by the considered prepulse.

To verify the robustness of the acceleration scheme against realistic laser contrast effects, we conducted a set of radiation hydrodynamics simulations using FLASH code \cite{FLASH} for the parameters described in Section III. Obtaining a set of density snapshots for 0.2 - 1 ns into the laser pedestal-CH target interaction, we initialized 2D PIC runs with these density snapshots and compared the resulting maximum proton energies at the end of 2D PIC runs. Figure \ref{fig:flash}a shows the electron density snapshot from the FLASH run for the case of $10^{-7}$ laser contrast and 1 ns into the simulation. We may see that while the target density departed from the initial channel structure location shown in dashed black lines, the overall structure of the target remains intact. Figure \ref{fig:flash}b reveals the effect of the laser pedestal on the maximum ion energies obtained in these simulations. We find that the presence of the pedestal with $\leq 400~ \rm ps$ duration and contrast no worse than $10^{-7}$ keeps the peak ion energy within the 75\% of the ideal case of no prepulse. Thus, we may conclude that the realistic laser contrast of moderate parameters does not reduce the efficiency of the acceleration mechanism. A set of FLASH+PIC runs with the uniform CH target were also considered, delivering significantly diminished peak ion energies (circle-dotted blue line in Figure \ref{fig:flash}b). A more detailed analysis of radiation hydrodynamics + PIC pipeline simulations is required for better matching experimental conditions of a particular laser facility, including realistic 3D geometry, target material, and oblique incidence. These questions are beyond the scope of this paper and will be addressed separately.

\begin{figure}
    \centering
    \includegraphics[width=\linewidth]{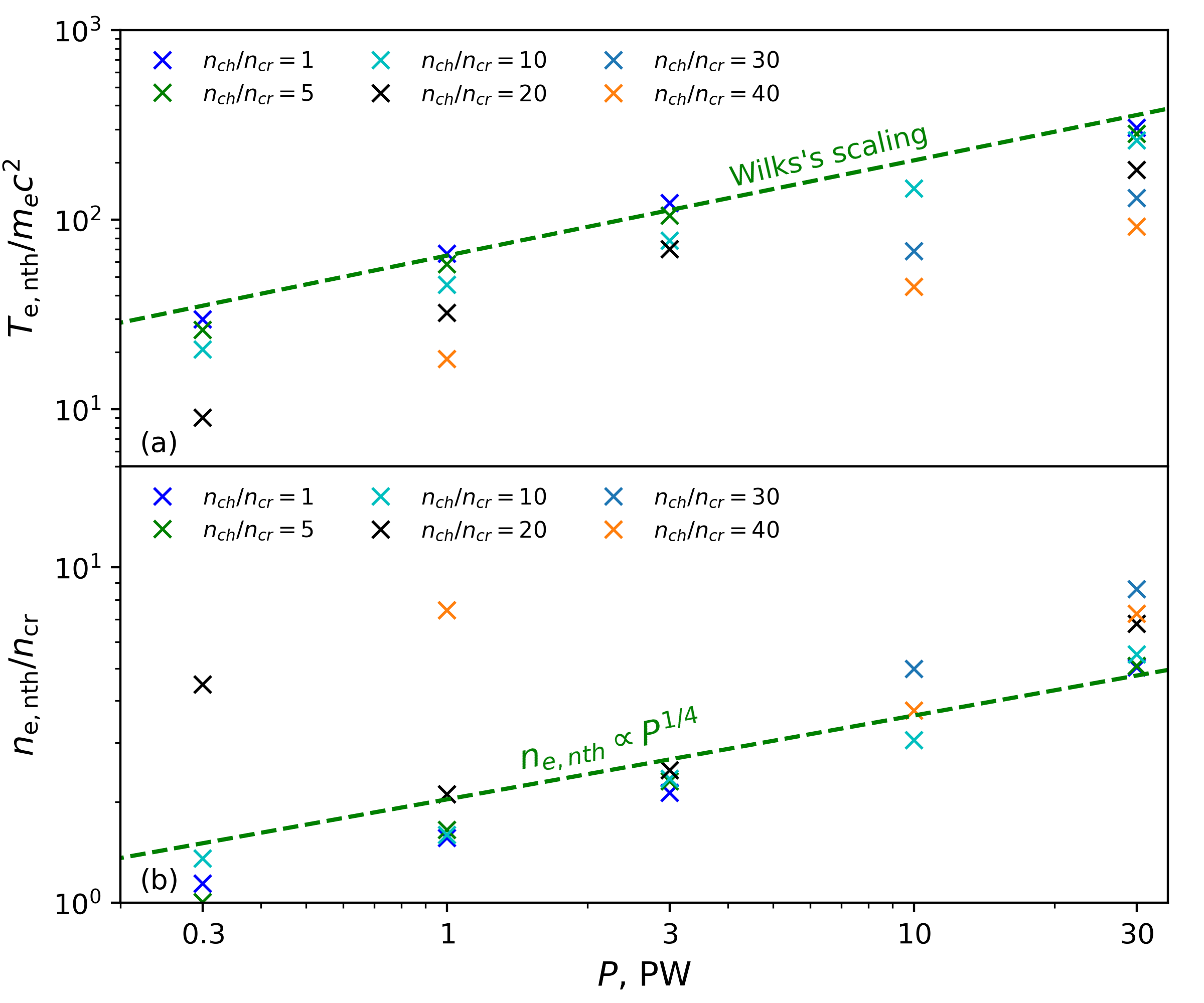}
    \caption{Scalings of (a) electron non-thermal temperature, $T_{\rm e,nth}$ and (b) density $n_{\rm e,nth}$ with laser power. Wilks' scaling and waveguide model (green dashed lines in (a) and (b), respectively) fairly explain the simulation results.}
    \label{fig:nth}
\end{figure}

\begin{figure*}
    \centering
    \includegraphics[width=\linewidth]{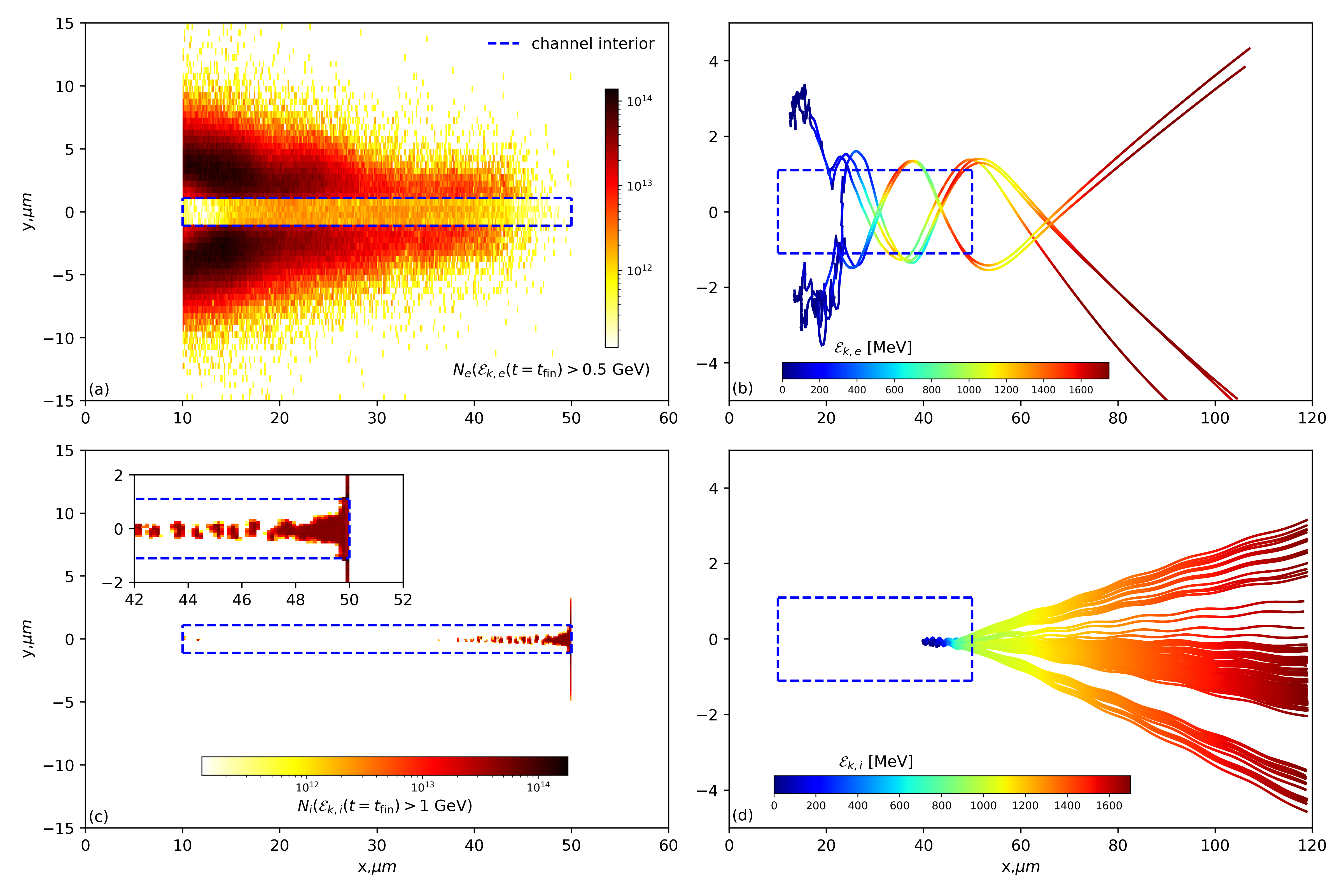}
    \caption{Histograms of original location of (a) fastest electrons (ones that end up with kinetic energy $>500$ MeV) and (c) fastest ions ($>1$ GeV); fast electron (b) and ion (d) tracks for the simulation with $P=10$ PW, $L_{\rm ch}=40 \mu \rm m$, and $n_{\rm ch}/n_{\rm cr}=20$. Blue dashed lines denote the initial location of the channel interior.}
    \label{fig:origin}
\end{figure*}

For a better understanding of the TNSA stage of ion acceleration, it is of interest to calculate the average density and temperature of hot electrons to the right from the rear end of the channel. Figure \ref{fig:nth} demonstrates the scaling of non-thermal electron population parameters with laser pulse power. Fig. \ref{fig:nth}a shows such dependence for the non-thermal temperature ($T_{\rm e,nth} \equiv \langle \mathcal{E}_e \rangle$) and compares it to the established scalings \citep{Zou2017}. Fig. \ref{fig:nth}b reveals calculations of average non-thermal electron density, $n_{\rm e,nth}$, and compares it against the waveguide model \citep{Zou2017}. Overall, the non-thermal temperature measured in all simulations is in fair agreement with Wilks' scaling \cite{SWILKS}, $T_{\rm e,nth}\propto \sqrt{1+a_0^2}-1 \approx a_0$. $n_{\rm e,nth}$ scaling is fairly captured by the waveguide model, although the best fit suggests a stronger dependence of non-thermal electron density on laser power. It may be explained by different optimal target parameters $L_{\rm ch}$ and $n_{\rm ch}$ for the considered range of laser pulse powers $P=0.3-30$ PW. Backtracking all electrons that end up with kinetic energies larger than 500 MeV for $P=10$ PW, $L_{\rm ch}=40 \mu \rm m$, $n_{\rm ch}=20 n_{\rm cr}$ run, we found that they mainly originate from the front side of the channel walls (see Figure \ref{fig:origin}a), specifically, from two lobes centered around $x= 15 \lambda,\, y = \pm 2-3 \lambda$, which is within the limits predicted by waveguide theory ($d_e = c/\omega_{pe} \approx \lambda \sqrt{a_0 n_{\rm cr}/n_{\rm wall}}/2 \pi \sim \lambda$). Simulations with smaller $P$ and/or $n_{\rm ch}$ lead to a similar conclusion. Figure 9b exemplifies a few trajectories of the fastest electrons in the simulation. They also originate from solid target walls and enter the oscillation cycle in the involved configuration of laser and background fields \cite{Mangles2005,Gong2020,Jirka2020}, effectively gaining energy at the channel exit. Finally, we calculated the relative role of channel filling and wall electrons in the TNSA accelerating field by comparing $\sqrt{n_{\rm e,nth}T_{\rm e,nth}}$ for each electron population. The fraction of channel filling contribution to the electrostatic field was found to be no more than $30 \%$, thus further suggesting a secondary role of channel filling.

It is also worth discussing where the fast ions are originated. In order to do so, we conducted (1) a simulation with $P = 10$ PW, $n_{\rm ch}/n_{\rm cr}=20$, $L_{\rm ch}= 40 \,{\rm \mu m}$ with a full tracking of ion trajectories and (2) a set of simulations with $P = 10$ PW, $L_{\rm ch}= 40 \,{\rm \mu m}$, varying $n_{\rm ch}/n_{\rm cr}$ from 0 to 40 and tagging channel filling and wall ions. The first run suggested that the fastest ions (ones that ended up with kinetic energy exceeding $1$ GeV) are predominantly from the rear end of the channel filling (see Figure \ref{fig:origin}c), while a small yet non-negligible fraction originated from the channel walls withing a few microns from the channel rear end. Analogous runs with the smaller filling density of $n_{\rm ch}/n_{\rm cr}=1$ or smaller laser pulse power $P=1$ PW are in qualitative agreement with this finding. The channel filling density scan with tagged particles revealed that the high-energy end of the ion spectrum (ions with 100 MeV or more) is comprised of both filling and wall ions for all runs, with wall ions dominating in $n_{\rm ch}/n_{\rm cr} \leq 10$ range and filling ions being abundant for $n_{\rm ch}/n_{\rm cr} \geq 20$, including the case of optimal channel filling density. Figure \ref{fig:origin}d shows a few tracks of fast ions that ended up with $\mathcal{E}_{k,i}\approx 1.5~ \rm GeV$ accelerated from the rear end of the target, again verifying the two-stage nature of the ion acceleration scheme: it acts both at the rear end of the target and further away from the sheath field.

\begin{figure}
    \centering
    \includegraphics[width=\linewidth]{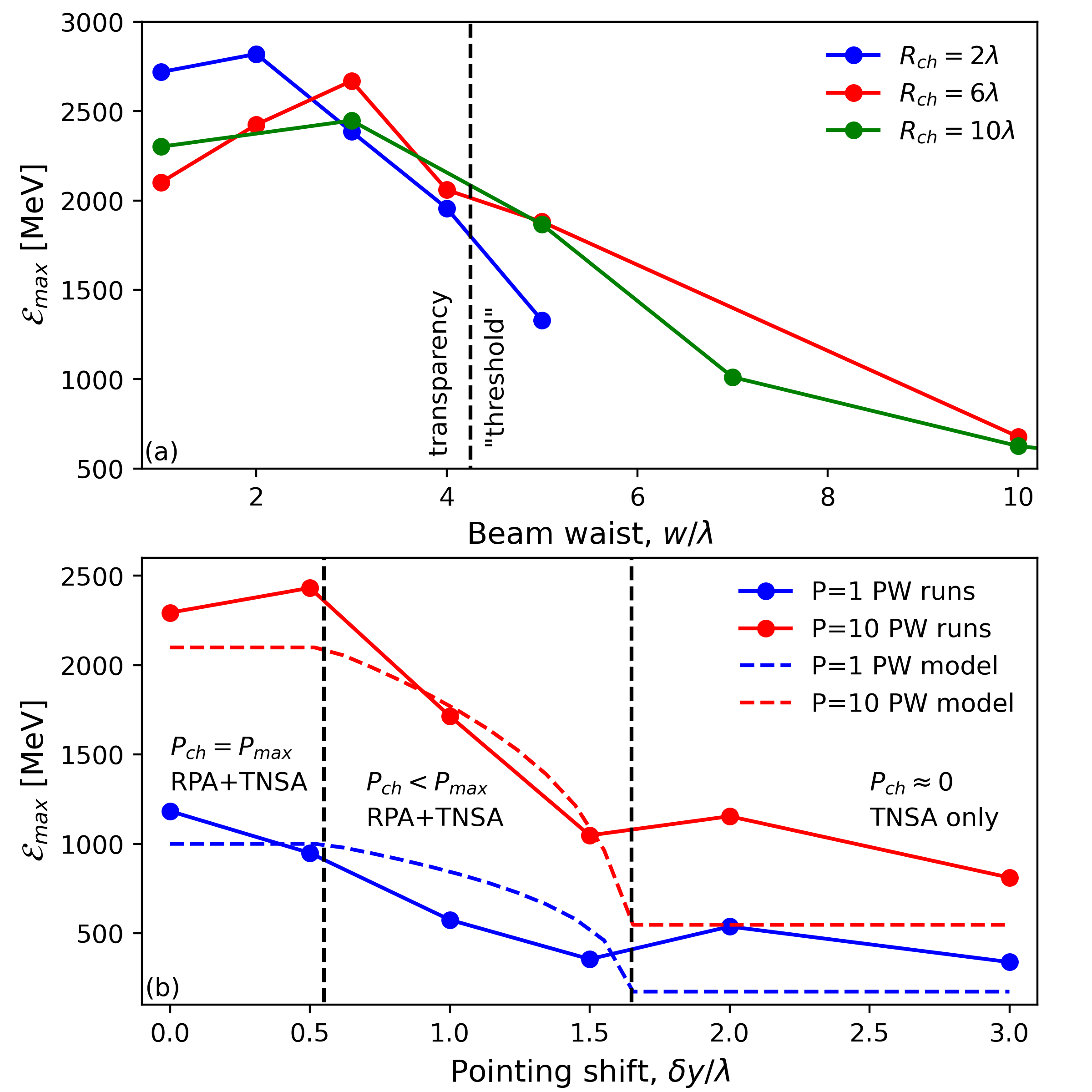}
    \caption{(a) Dependence of maximum ion energy on beam waist for different channel radii $R_{\rm ch}=2-10 \mu \rm m$ for $P=10$ PW, $L_{\rm ch}=40 \mu \rm m$, $n_{\rm ch}=10 n_{\rm cr}$ simulations. Vertical dashed line denotes an approximate threshold for relativistic transparency of the channel to hold. (b) maximum ion energy dependence on laser axis shift $\delta y$ for $P=1$ PW (blue) and $10$ PW (red). Dashed lines represent the predictions of theoretical model.}
    \label{fig:focus}
\end{figure}

As one may see from Eqn.~\ref{eqn:maxenergy}, the contribution from channel radius and beam waist to the maximum attainable ion energy is connected to the energy transfer efficiency of the laser pulse with the constant beam waist to the channel filling electrons under the assumption that the channel does not significantly evolve. In reality, however, these assumptions do not hold, as we see from our runs with varying initial beam waists and channel radii. When channel radius, $R_{\rm ch}$, is significantly larger than beam waist at focus, $w$, laser pulses experience transverse filamentation and hosing instabilities \cite{Naumova2001}, leading to the reduced ion acceleration efficiency. At the same time, for $w>R_{\rm ch}$, the laser energy may be partially scattered from the channel entrance, decreasing the acceleration efficiency. Figure \ref{fig:focus}a presents our scan on laser beam waist for $P=10$ PW laser pulse and $n_{\rm ch}=20 n_{\rm cr}$ target with $R_{\rm ch} = 2, 6, 10 \lambda$. A sweet spot is found around $w_{\rm opt} \approx~ 1-3 \mu \rm m$ for all channel radii considered. This is below the radius of self-channeling laser pulse in uniform plasma given by $R_{\rm sc}/\lambda = 1/\pi (n/n_{\rm cr})^{1/3}(27 P/P_{\rm cr})^{1/6} \approx 14$ \citep{SSBulanov2010}. To understand why maximum ion energy drops for $w>4 \lambda$, we recall that the considered two-stage acceleration mechanism requires relativistic transparency of the channel for efficient laser-target coupling and ion acceleration via RPA. Relativistic transparency regime is realized when $n_{e}\ll \langle \gamma_e \rangle n_{\rm cr}$. If we recall that $\langle \gamma_e \rangle \approx a_0$ and expressing $a_0$ as the function of laser pulse power $P$ and beam waist $w$, $a_0  = 0.85 \sqrt{P/(I_0 w^2)}$, we get the threshold value of $w_{t} \approx 4.25 \mu \rm m$, above which the relativistic transparency is violated, and the two-stage mechanism transitions to classic TNSA, thus decreasing the maximum ion energy. In other words, the channel radius constraints are quite relaxed, requiring only $R_{\rm ch}>w$ for optimal acceleration. The laser focusing, however, should be narrow enough to trigger relativistic transparency within the channel, i.e. $w/\lambda < w_t/\lambda \approx \alpha \sqrt{10^5 P[\rm PW]}/(n_e/n_{\rm cr})$, where $\alpha \ll 1$ is a dimensionless factor controlling the relativistic transparency condition, $n_e = \alpha a_0 n_{\rm cr}$, which was assumed to be equal to 0.1 in our calculations. For the smallest power considered, $P=0.3$ PW, it requires a beam waist of $w/\lambda < 1.73$, which is within reach for modern laser systems \cite{Yoon2021}.

Transverse laser pointing stability is also governed by a similar physical process. When laser beam completely misses the channel cross-section, i.e. when the absolute value of the laser axis pointing shift, $\delta y$, is equal or larger than $R_{\rm ch}+w/2$, TNSA mechanism is realized with an increased hot electron population due to presence of the channel. When the laser beam spot is within the channel cross-section, i.e. when $|\delta y| \leq R_{\rm ch}-w/2$, there should be no changes in maximum ion energy according to the proposed theoretical model. For the pointing shifts in between these two values, the maximum ion energy is obtained from the two-stage mechanism with the effectively reduced laser pulse power, $P_{\rm ch} \approx P (0.5+(R_{\rm ch}-\delta y)/w)^{N-1}$, where $N$ is the dimensionality of the problem. The maximum ion energy is calculated from our theoretical model for $P_{\rm ch}>0$ and from classical scaling $\mathcal{E}_{\rm max} = 173 \rm ~ MeV \sqrt{P[\rm PW]}$ for $P_{\rm ch} =0$ \cite{Esirkepov2014}. Figure \ref{fig:focus}b compares our simulation results for pointing stability scan for $R_{\rm ch}=1 ~\mu \rm m$ and $w=1.1 ~ \mu \rm m$ with our interpretation. The agreement is satisfactory, with the largest discrepancy appearing in the case of TNSA-only accelerated ions. The final requirement on maximum pointing shift is $|\delta y|\leq \delta y_{\rm max}= R_{\rm ch}-w/2$.

\begin{figure}
    \centering
    \includegraphics[width=\linewidth]{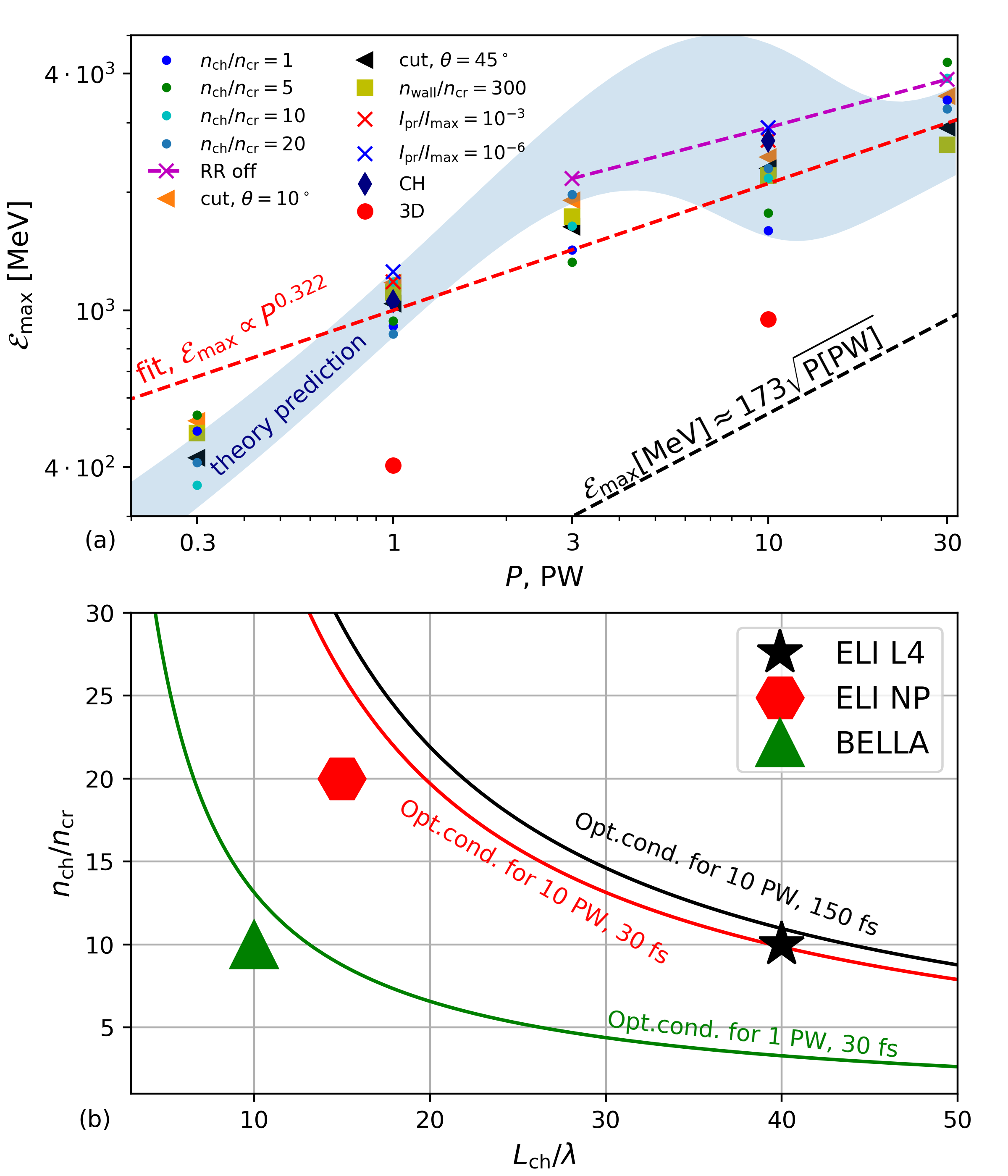}
    \caption{(a) Maximum energy scaling with power. Different markers represent different channel densities; the dot-dashed line corresponds to the simulations with RR off. We also plot typical energy scalings - classical theoretical scaling \cite{Esirkepov2014} (black dashed line). The fit of maximum ion energies observed in our runs (red dashed line) is also shown. (b) Optimal target conditions for ELI L4, ELI-NP, and BELLA lasers from 2D PIC simulations (markers) and theory predictions (Eqn.~\ref{eqn:optcond}), solid lines).}
    \label{fig:scaling}
\end{figure}

To understand the possible range of applicability of the considered acceleration scheme, it is of interest to check such a target on the maximum energy scaling with laser pulse power. Figure \ref{fig:scaling} summarizes the whole set of our simulations. It turns out that the scaling derived in Section II (shaded region) is in approximate agreement with simulations. Maximum proton energies are well above the usual theoretical scaling for maximum proton energy (black dashed line) \cite{Esirkepov2014}. The universal fitting formula for maximum ion energy scaling may be written as:

\begin{equation}
    \mathcal{E}_{\rm max} =  \left(\frac{\rm P}{\rm 1\,PW} \right)^{0.322} \, {\rm GeV}.
    \label{eqn:scaling}
\end{equation}

\noindent Runs with RR off demonstrate a factor of a few advantage in terms of maximum ion energy in contrast to RR on simulations (cross-dashed magenta line in Fig.\ref{fig:scaling}), which is expected as the additional laser energy losses given by Eqn.~\ref{eqn:EnRR} diminish both TNSA and RPA efficiency.

We also considered a cut on the front side of the target, having a wedge with $\theta =10^\circ$ and ${ 45^\circ}$ angle on its front. It is evident that such a cut does not suppress ion acceleration (orange and black triangles in Fig.\ref{fig:scaling}), and may be helpful for experimental realization of the proposed acceleration scheme by avoiding hazardous laser backreflection \cite{Snyder2019,Bailly2020}. A slight decrease in maximum ion energy for the cut of $45^\circ$ may be interpreted on a basis of Figure \ref{fig:origin}a, where the cut of a target front may effectively decrease the efficiency of hot electron generation by suppressing $n_{\rm e,nth}$.

The additional runs with the increased wall density ($n_{\rm wall}/n_{\rm cr}=300$) do not differ much from our main set of simulations with $n_{\rm wall}/n_{\rm cr}=100$, since the dependence of channel wall skin depth on the solid target density is pretty weak, $d_e \propto n_{\rm wall}^{-1/2}$. As a result, fast electron generation is not affected, leading to similar values of $\mathcal{E}_{\rm max}$ (squares in Fig.\ref{fig:scaling}). Likewise, considering a realistic CH target with $n_{\rm wall}/n_{\rm cr}=300$ and $n_{\rm ch}/n_{\rm cr}=20$ \cite{Rinderknecht2021}, we observe the same level of maximum proton energies (diamond markers in Fig.\ref{fig:scaling}a) and ion acceleration mechanism.

Field ionization was also included in a separate series of 2D PIC runs. Both picosecond prepulse and driver pulse are capable of fully ionizing the part of the channel responsible for ion acceleration, and the maximum ion energies are not affected for all considered laser pulse powers. The presence of picosecond prepulses did not change the maximum attainable ion energies, as discussed earlier in the paper. 

Finally, we conducted a series of 3D runs, which showed relatively smaller maximum proton energies for $P=1 \& 10$ PW laser pulses in comparison to our 2D runs, possibly, due to weaker hot electron retention and stronger constraints on the transparency of the accelerated foil than in 2D \cite{Psikal2021}. Still, the main features of the acceleration mechanism, namely, rapid ion acceleration from the channel rear end at the time of laser pulse exiting the channel and presence of quasi-monoenergetic structure in the ion energy spectrum were verified, in agreement with \cite{Arefiev2018}.

\section{Summary and Discussion}

In this paper, we considered laser ion acceleration from the micron-scale channels filled with relativistically transparent plasma. We derived an optimal set of parameters for such acceleration and obtained a model to predict the ion energy gain with time. These considerations were checked against 2D PIC simulations and are in fair agreement with them. A few experimentally relevant physical effects were also addressed. The main results may be listed as follows:

\begin{itemize}
    \item The acceleration is interpreted as a combination of TNSA and RPA, illustrated by Figures \ref{fig:scheme} and \ref{fig:phspace}.
    \item Quasi-monoenergetic features in the wide high-energy proton spectra are observed, as shown in Figure \ref{fig:spec}. The time of quasi-monoenergetic structure development coincides with the time of laser pulse exiting the channel.
    \item A theoretical model is developed on a basis of \cite{SSBulanov2010,SVB2014,SWILKS}. Optimal interaction conditions are given by Eqn.~\ref{eqn:optcond}, in approximate agreement with 2D PIC scans on channel length $L_{\rm ch}$ and channel filling density $n_{\rm ch}$, depicted by Figures \ref{fig:emax_vs_t_lch} and \ref{fig:emax_vs_nch}.
    \item The role of laser contrast is investigated, both indirectly, via the analysis of the role of channel filling electrons, and directly, with picosecond prepulse PIC simulations and auxiliary radiation hydrodynamics simulations of nanosecond pedestal coupled with PIC. Figures \ref{fig:emax_vs_nch} and \ref{fig:filling} illustrate the relaxed conditions on the channel filling density, while Figures 8 and 9 discuss the role of channel filling on electron heating.
    \item Channel radius requirements are shown to be a non-restrictive factor in the discussed acceleration scheme, see Figure \ref{fig:focus}a, which is beneficial for the currently available channel targets \cite{Rinderknecht2021}.
    \item The limiting factor for beam waist was found to be a relativistic transparency threshold, $w<w_t$. As long as it is smaller than the channel radius, $w<R_{\rm ch}$, the maximum ion energies are unaffected.
    \item The acceleration mechanism is shown to be robust to moderate perturbations in laser pointing, see Figure \ref{fig:focus}b, in fair agreement with the model.
    \item Oblique incidence (i.e. a cut at the front surface of the target) and field ionization were shown to be not significant limitations for maximum ion energies.
    \item In our 2D PIC scan, we observed GeV-scale protons accelerated by PW-scale laser pulses with approximate energy scaling $\mathcal{E}_{\rm max} \propto P^{0.322}$, seen in Figure \ref{fig:scaling}.
    \item Finally, three-dimensional simulations verified the primary features of the acceleration scheme, though the maximum ion energies are less than in analogous 2D cases.
\end{itemize}

One may argue that the suboptimal power scaling questions the applicability of structured targets. While power scaling was found to be quite shallow (Eqn.~\ref{eqn:scaling}), for $P\leq 1 \rm PW$ the acceleration mechanism is competitive with other mechanisms \cite{SSBulanov2010,Arefiev2018} in terms of maximum ion energies. Moreover, an additional analysis suggested that the considered mechanism possesses a high laser-to-proton energy conversion efficiency of no less than 15\%, promising a high volumetric charge of fast ion beam \cite{Arefiev2018}. This feature of the discussed acceleration scheme may be beneficial for the fast ignition concept in inertial confinement fusion \cite{MROTH,Honrubia2009}.

In comparison to uniform near-critical density targets (or, for the channels with channel radius significantly exceeding beam waist, $R_{\rm ch}\gg w$), channel targets provide better pulse guiding and larger counts of fast protons \cite{Arefiev2018}. Auxiliary simulations with uniform near-critical target with $n_{\rm e} = 1-40 n_{\rm cr}$ show that the fast ion population is an order of magnitude smaller than for channel target, along with a notable laser pulse hosing, detrimental for the resulting ion source angular distribution.

Quasi-static magnetic fields produced by micron-scale channel targets are remarkable - they exist on a picosecond scale, and demonstrate maximum values $B_{\rm max}^{\rm QS} \approx 110~ {\rm kT} \cdot (P[\rm PW])^{0.2}$ even after the pulse exits the channel, closely reaching the megatesla-scale magnetic fields predicted in microtube implosions \cite{Murakami2020}. These fields are known to significantly modify the electron motion inside the channel, allowing for a steady energy gain \cite{Wang2020a}, and may provide a platform for experiments with $\gamma$-ray generation and pair production \cite{Stark2016,Jansen2018,Rinderknecht2021}. The obtained $B_{\rm max}^{\rm QS}$ values are smaller than $B_{\rm max}^{\rm QS} \approx 550~ {\rm kT} (P[\rm PW])^{0.5}$ suggested by Eqns. 2\&3 in \cite{Arefiev2018} due to the difference in the magnetic field measurement methodology.

It is worth noting that Magnetic Vortex Acceleration (MVA) mechanism may also contribute to the ion acceleration at the rear side of the target. Indeed, as we observe a strong quasi-static magnetic field forming inside the channel, we may expect the dipole structure to expand out of the channel exit, thus maintaining the charge separation and corresponding sheath field. However, since the considered channel length is smaller than the optimal pulse dissipation length for MVA obtained by equating Eqns.~\ref{eqn:Elas} and \ref{eqn:Eele} \cite{SSBulanov2010}, and the solid wall density preventing the magnetic vortex expansion, we believe that the MVA is suppressed in our acceleration scheme.

When choosing the parameters for the simulations, we aimed at those that will soon be available on ELI-Beamlines L4 ATON laser \cite{ELIBL}. Based on our analytical model, we may envision an efficient application of the discussed acceleration scheme with laser parameters of ELI-NP \cite{ELINP2}, Apollon \cite{APOLLON}, J-KAREN-P \cite{KIRIYAMA2018}, and BELLA \cite{Leemans2013} as well. Figure \ref{fig:scaling}b shows optimal structured target conditions for these lasers obtained through auxiliary 2D PIC scans and theoretically predicted optimal regime given by Eqn.~\ref{eqn:optcond}. The agreement between them is fair, though the maximum proton energy for 1 PW, 30fs laser pulse is significantly suppressed, being no more than 600 MeV.

Finally, let us discuss how the considered target compares to the microstructure targets produced today. In \cite{Rinderknecht2021}, a very similar type of target was considered, with primary differences being channel dimensions and material. Our results suggest that the maximum ion energy will be suppressed for the realistic channels of $L_{\rm ch} \sim 100~\mu \rm m$, giving a preference for channels of approximately half of that size, as seen in Figures 4c,d. Scans on channel radius (Figure \ref{fig:focus}a) and pointing stability (Figure \ref{fig:focus}b) predict a promising scaling to realistic parameters $R_{\rm ch}=6 ~\mu \rm m$ and $\sqrt{ \delta y^2} = 5 ~ \mu \rm m$, sustaining the acceleration efficiency. Our simulations with increased solid wall density ($n_{\rm e, wall}=300 n_{\rm cr}$) and high-Z runs for polystyrene target (radiation hydrodynamics + 2D PIC simulations) and CH target (2D PIC simulations; diamonds in Figure \ref{fig:scaling}a) suggest that the considered laser ion acceleration scheme will be applicable for the Kapton substrate-CH foam filling targets as well, in agreement with \cite{Arefiev2018}.

The results obtained in the paper show that the considered laser ion acceleration scheme is robust against moderate variations in laser and target parameters, thus making it a viable candidate for experimental implementation.

\section*{Acknowledgements} 
This work was supported by NNSA DE-NA0003871, DE-SC0021248, and AFOSR FA9550-15-1-0391 and by the project 
High Field Initiative (CZ.02.1.01/0.0/0.0/15 003/0000449) from European Regional Development Fund.
 The EPOCH code was developed as part of the UK EPSRC funded projects EP$/$G054940$/$1. The software used in this work was developed in part by the DOE NNSA- and DOE Office of Science-supported Flash Center for Computational Science at the University of Chicago and the University of Rochester. The simulations presented in this article were performed on computational resources managed and supported by Princeton Research Computing at Princeton University. K.V.L. is thankful to Alexey Arefiev for fruitful discussions.

\end{document}